\documentstyle[12pt,epsf]{article}
\newcommand{\beq}[1]{\begin{equation}\label{#1}}
\newcommand{\eeq}{\end{equation}}
\newcommand{\beqar}[1]{\begin{eqnarray}\label{#1}}
\newcommand{\eeqar}{\end{eqnarray}}
\newcommand{\cs}{{\cal S}}
\newcommand{\al}{\alpha}
\newcommand{\be}{\beta}  
\newcommand{\ep}{\varepsilon}
\newcommand{\ga}{\gamma}
\newcommand{\de}{\delta}

\newcommand{\ka}{\kappa}
\newcommand{\la}{\lambda}

\newcommand{\si}{\sigma}

\newcommand{\om}{\omega}

\newcommand{\ze}{\zeta}

\begin{document}
\vspace*{-2cm}
\hfill UFTP 416 preprint/1996 

\hfill NTZ 26/96

\hfill TUM/T39-96-16
\vspace{1.5cm}
\begin{center}
{\LARGE \bf Small $x$ behaviour of the chirally-odd parton 
    distribution $h_1(x,Q^2)$.}\\[2mm]
\vspace{1cm}
         {\large R.~Kirschner
} \\
\vspace{.5cm}
Institut f\"ur Theoretische Physik and Naturwissenschaftlich-Theoretisches
Zentrum,  \\
Universit\"at Leipzig, 
Augustusplatz, D-04109~Leipzig, Germany \\
\vspace{1cm}
  {\large L.~Mankiewicz}\footnote{On leave of absence from N. Copernicus 
Astronomical Center, Polish Academy of Sciences, Warsaw, Poland.}
\\
\vspace{.5cm}
 Institut f\"ur Theoretische Physik, Technische Universit\"at M\"unchen, \\
  D-85747   M\"unchen, Germany \\
\vspace{1cm}
  {\large A.~Sch\"afer and L.~Szymanowski}\footnote{On leave of absence from
 Soltan Institute for Nuclear Studies, Warsaw, Poland.}
\\
\vspace{.5cm}
Institut f\"ur Theoretische Physik, J.~W.~Goethe
Universit\"at Frankfurt,
\\
 Postfach 11 19 32, D-60054~Frankfurt am Main, Germany
 
\end{center}

\vspace{1.5cm}
\centerline{\bf Abstract:}

\vspace{.5cm}
\noindent The small $x$ behaviour of the 
 structure function $h_1(x,Q^2)$ is 
studied within the leading logarithmic approximation of  perturbative QCD.
There are two contributions relevant at small $x$. The leading one behaves
like  $(\frac{1}{x})^0$ i.e. it is just 
 a constant in this limit. 
The second contribution, suppressed by one
power of $x$, includes the terms summed by the GLAP equation. Thus for
$h_1(x,Q^2)$ the GLAP asymptotics and Regge asymptotics are completely
different, making $h_1(x,Q^2)$ quite an interesting quantity for the study of
small $x$ physics. 

\vspace*{1cm}
\centerline{\it Dedicated to Professor Wojciech Kr\'olikowski.}
 
\newpage

\section{Introduction.}

The chirally-odd structure function $h_1(x,Q^2)$ appeared  
first in studies 
by Ralston and Soper \cite{RS}
of the spin asymmetries   in
 Drell-Yan processes (DY) for transversely polarized beams and targets.
Its appearence is due to the fact that in DY processes
the chiralities of quark lines originating in the same hadron are 
not correlated and amplitudes with different quark chiralities interfere.
 This is not possible for deep-inelastic
scattering (DIS) where the chiralities of the quark lines have to be the same.
The physical interpretation of $h_1(x,Q^2)$ structure function was given
by Jaffe and Ji \cite{JaffeJi91}: it measures the transversity asymmetry i.e. 
the difference of the probabilities 
to find a quark with spin polarized along the spin direction of a
transversely  polarized
nucleon and  respectively opposite to it.

Recently we are faced with  the revival of interest in studies of 
polarized proton-nucleon collisions, both theoretically and experimentally. 
Several experiments at polarized hadronic colliders
are planned for the near future. The physical
motivation for those experiments is reviewed 
in Refs.~\cite{exp} and recently by Jaffe in Ref.\cite{Jaffe96}. 
In particular, the RHIC spin collaboration at BNL 
 will provide in the near future a first measurement of $h_1(x,Q^2)$. 
Also the COMPASS collaboration at CERN plans to 
study this structure function.

The results of theoretical studies for $h_1(x,Q^2)$ are very limited so far.
 Because
its decoupling from DIS this structure function was not an object of
many investigations. A summary of results can be found in 
Ref.~\cite{JaffeErice}. Below we would like to mention only 
some of the results which 
are relevant for the present work. The GLAP evolution
equations for $h_1(x,Q^2)$ were studied by Artru and Mekhfi 
\cite{ArtruMekhfi}. The anomalous dimensions
corresponding to $h_1(x,Q^2)$ appeared already earlier in the Appendix of
Ref.~\cite{BuFKL}. Collins has proved the factorization of hard and 
soft parts in  DY processes, involving $h_1(x,Q^2)$ \cite{Collins}.
Recently Ioffe and Khodjamirian have calculated  $h_1(x,Q^2)$ by means of the
QCD sum rules \cite{IK}. 

The longitudinal spin effects and the transverse spin effects appear on 
equal footing in perturbative QCD.
As both polarized structure functions $g_1(x,Q^2)$, measuring the helicity
distribution and $h_1(x,Q^2)$, measuring the transversity distribution,
contribute at the twist-two level to hard scattering processes,  it is
especially interesting to investigate their similarities and differences. In the 
non-relativistic limit these functions coincide but already in the bag model
they are different \cite{JaffeJi91}. 
 
We want to concentrate on the properties of polarized structure functions
$g_1(x,Q^2)$ and $h_1(x,Q^2)$ in the kinematical region of small Bj\"orken
variable $x$. While the small-$x$ behaviour of $g_1(x,Q^2)$
is  the subject of recent studies by Bartels, Ermolaev and Ryskin
\cite{BER}, we discuss in  
this  paper  the small-$x$ behaviour of $h_1(x,Q^2)$.

 Jaffe and Ji define $h_1(x,Q^2)$ as Fourier transform of the
matrix element of the following bilocal quark operator 
on the light-cone \cite{JaffeJi91}
\beq{JJh}
h_1(x,Q^2) \cs_{\perp \mu} = \frac{i}{2} \int_{-\infty}^{\infty}
\frac{d\la}{2\pi} e^{i\la x} <p,\cs_\perp| \bar{\psi}(0) \sigma_{\mu\nu}n^\nu
\ga_5 \psi(\la n)|p,\cs_\perp> \;\;,
\eeq

where $x = Q^2/2p\cdot q$, $Q^2 = - q^2$, $n^\mu$ is the light-cone vector, 
$p^\mu$ and $\cs_{\perp}^{ \mu}$ are the momentum and the transverse part of 
the spin vector $\cs$ ( $\cs = (\cs\cdot n)p + (\cs\cdot p)n + \cs_\perp \,$).

As was shown by Ioffe and Khodjamirian the structure function
$h_1(x,Q^2)$ can also be related to a forward scattering matrix element between 
proton states  of an operator containing an axial-vector
and a  scalar current  \cite{IK}
\beqar{IKh}
T_\mu(p,q,\cs_\perp) &=& i \int d^4x e^{iqx} <p,\cs_\perp| 
\frac{1}{2} T{\big (} j_{5\mu}(x) j(0) + j(x) j_{5\mu}(0){\big )}|p,\cs_\perp>
  \nonumber \\
    &=& {\tilde h}_1(x,Q^2) \cs_{\perp \mu}
\eeqar
where $j_{5\mu}(x) = \bar{\psi}(x)\ga_5\ga_\mu\psi(x), j(x) =
\bar{\psi}(x)\psi(x)$ and in the second line of the Eq.(\ref{IKh}) only
terms depending on the spin polarization vector $\cs$ are
retained. Then, up to higher twist terms one gets
\beq{IKh1}
   h_1(x,Q^2) = - \frac{1}{\pi} Im \;{\tilde h}_1(x,Q^2).
\eeq

The perturbative contribution to $h_1(x,Q^2)$ involves the $t$-channel
exchange of a quark and an antiquark of opposite chirality (parallel helicity).
Amplitudes with such an exchange behave at large energy like $s^0$ plus
logarithmic corrections \cite{Asimov}. This implies that for small $x$
 $h_1(x,Q^2)$ is roughly like $(\frac{1}{x})^0$ i.e like a constant. 
On the other hand the
one-loop parton splitting function is proportional to $x \approx 
\frac{Q^2}{s}$ at $x \rightarrow
0$. Therefore  GLAP evolution \cite{GLAP} induces an asymptotics 
proportional to $x$,
provided the input distribution has not a stronger asymptotics
 at $x \rightarrow 0$. The
expectations from these simple arguments are confirmed by the following
detailed analysis of the leading logarithm perturbative contributions.

The structure function $h_1(x,Q^2)$ contributes to the DY
hadronic tensor ($p_A \, p_B \to \mu^+ \, \mu^- + X$) \cite{JaffeJi91} as
\beqar{h1def}
W_{DY}^{\mu\nu} & \sim & \Bigg[ (\cs_{A\perp}\cdot \cs_{B\perp}) (p_A^\mu p_B^\nu
+ p_A^\nu p_A^\mu - g^{\mu\nu} p_A \cdot p_B)  \nonumber \\
 & + & (\cs_{A\perp}^\mu \cs_{B\perp}^\nu + \cs_{A\perp}^\nu
\cs_{B\perp}^\mu)
p_A \cdot p_B \Bigg] \sum_q e_q^2 h_1^q(x) h_1^{\bar{q}}(y) \,
\eeqar
where the sum runs over all quarks $q$ and antiquarks ${\bar q}$.
Because of the symmetry in $(\mu\nu)$ the signatures in the exchange
channels to the hadron A and B are the same. Unlike in the case of
$g_1$ in DIS, here both signatures $h_1^{\pm} =
h_1^q \pm h_1^{\bar q}$ contribute. A definite signature contribution
can be extracted by combining the data for DY production in
proton-proton and proton-antiproton collisions.

The definition (\ref{IKh}) leads to a positive signature amplitude. In
\cite{IK}
also another definition is mentioned, which leads to a negative
signature amplitude. In the following we shall investigate the
perturbative Regge asymptotics in the positive signature channel,
starting from the formula (\ref{IKh}).

In section 2 we study the leading Regge asymptotics induced by the exchange
of quark and antiquark. We rely on ideas related to the well known BFKL
perturbative Pomeron \cite{BFKL} and the multi-Regge effective action
\cite{KLS}. The leading logarithm approximation applied here does not
include the coupling constant renormalization. The coupling constant in the
 formulas has to be considered as a fixed, determined by the typical
 transverse momentum scale.

In section 3 we consider the double logarithmic contributions including the
ones described by  GLAP evolution at small $x$. Here we rely on the
method of separation of soft particle \cite{KL} which has been applied
recently in the studies of $g_1(x,Q^2)$ \cite{BER}.

\section{The leading log Regge asymptotics.}

\setcounter{equation}{0}
\subsection{Lowest order contribution.}

We calculate the lowest order contribution to $h_1(x,Q^2)$ first. For this
purpose
we adopt the parton model viewpoint and identify the proton with the on
mass-shell quark.
 Thus the lowest order
contribution to $T_\mu(p,q,\cs_\perp)$ (Eq.(\ref{IKh}) 
is given by four graphs  in Fig.~1. We obtain
\beq{bornT}
   T^{(0)}_\mu(p,q,\cs_\perp) = \cs_{\perp\mu} \{ \frac{s}{s - Q^2 + i\ep} +
\frac{-s}{-s - Q^2 +i\ep} \} 
\eeq
where $s = 2p\cdot q$. The $s$-channel immaginary part of Eq.(\ref{bornT}) leads 
to the usual lowest order result  for the structure function
in the parton model
\beq{bornh}
   h^{(0)}_1(x) = \de(1 - x).
\eeq

From Eq.(\ref{bornT}) we learned that $h_1(x,Q^2)$ defined via Eqs.(\ref{IKh})
and (\ref{IKh1}) is related to positive signature exchange unlike
$g_1(x,Q^2)$ which has negative signature \cite{BER}. 
The signature is relevant for the Regge asymptotics,
i.e. the small $x$ behaviour.

The leading log Regge asymptotics is calculated from graphs of the type
Fig.~2. Those are the effective ladder diagrams in which the reggeized
 quarks of opposite chiralities propagate on both sides 
 and interact through gluon exchange.

As a first step we consider the cut-vertex $\Phi^{(0)}$ coupling the exchanged
fermions to the external currents (see Fig.~2)
\beq{extcur}
          \ga_5 \ga_\mu ( \hat{k}_1 + \hat{q} ) 
              + ( \hat{k}_1 + \hat{q} ) \ga_5 \ga_\mu 
\eeq
where $\hat{k} = \ga^\mu k_\mu$.

Only the transverse components of the currents  are of interest, 
because of the projection
with the transverse polarization vector $\cs_{\perp \mu}$
(Eq.(\ref{IKh})). In the Regge
asymptotics the contributions of the momentum $k_1$ are negligible.

 It is convenient to adopt notations using light-cone components for the
longitudinal parts and complex numbers for the transverse parts of all
vectors, e.g.
\beqar{compnot}
k^\mu &=& \frac{2}{\sqrt{s}} ( q_1^\mu k_- + p^\mu k_+ ) + k^\mu_\perp \;\;,
\hspace{1cm} \ka = k^1_\perp + i k^2_\perp  \nonumber \\
q_1^\mu &=& q^\mu - \frac{q^2}{s}p^\mu \;\;.  
\eeqar
We consider the following projectors for  Dirac spinors
\beq{project}
P_+ = -\frac{1}{4}\ga \ga^* \;\;,\;\;\;\;P_- = -\frac{1}{4}\ga^*
\ga\;\;,\;\;\;\; \Pi_+ = \frac{1}{4}\ga_-\ga_+ \;\;,\;\;\;\; \Pi_- =
\frac{1}{4}\ga_+\ga_- \;\;.
\eeq 
Notice that $\ga_5 = i\ga_0\ga_1\ga_2\ga_3$ can be written as
\beq{g5}
    \ga_5 = (\Pi_+ - \Pi_-)(P_+ - P_-)
\eeq
and that spinors in the subspaces $\Pi_-P_+$ and $\Pi_-P_-$ 
(or in subspaces $\Pi_+P_-$ and $\Pi_+P_+$ ) correspond to
opposite chiralities.

For definitness we consider below in this section that combination of
amplitudes $T_\mu(p,s,\cs_\perp)$ (Eq.(\ref{IKh})) which couples to external
currents (\ref{extcur}) with the matrix $\ga^*$ 
\beq{T*}
T^*(p,s,\cs_\perp) \equiv - (T_1(p,s,\cs_\perp) - i T_2(p,s,\cs_\perp)) = 
\cs^* \tilde{h}_1(x,Q^2)\;\;.
\eeq 
Thus with  the notation (\ref{compnot}) we have
\beq{extcurrent}
 \ga_5 \ga^* ( \hat{k}_1 + \hat{q} ) + ( \hat{k}_1 + \hat{q} ) \ga_5
             \ga^* = q_-\ga^*\ga_+ + q_+ \ga_-\ga^* \;\;.
\eeq
The second term can be omitted since $q_+$ is proportional to $x =
\frac{Q^2}{s}$. We are left with the first term in Eq.(\ref{extcurrent}) as
the coupling of the external currents with the exchanged fermions.

\subsection{Reggeon interaction.} 

The leading logarithmic Regge asymptotics is determined by the exchange of
a reggeized quark and antiquark of opposite chirality. A typical
contribution is given by the graph in Fig.~2. The reggeons interact by gluon
exchange. The gluons are coupled to the exchanged quarks by the effective
vertex (Fig.~(3a)), which includes besides the original QCD vertex
contributions from bremsstrahlung production \cite{FS}, \cite{KLS}
\beq{effvertex}
\hat{V}^\mu(k_i,k_{i+1}) = \ga^\mu_\perp - 
\frac{\hat{k}_{i\perp}}{k_i\cdot p}\;p^\mu
- \frac{\hat{k}_{i+1 \perp}}{k_{i+1}\cdot q_1} \;q^\mu_1 \;\;.
\eeq

In Eq.(\ref{effvertex}) the colour group generators are omitted since they play
in the following a passive role.

The interaction kernel (Fig.~3b) is obtained by contracting two effective
vertices assuming the exchanged gluon to be on mass-shell.  We restrict
ourselves to  vanishing momentum transfer and we get
\beqar{VV}
\hat{V}^\mu(k_i,k_{i+1}) \otimes \hat{V}_\mu(k_{i+1},k_i) &=& 
  - \frac{1}{2}( \ga \otimes \ga^* + \ga^* \otimes \ga ) \frac{|\ka_i|^2 +
|\ka_{i+1}|^2}{|\ka_i - \ka_{i+1}|^2} \nonumber \\
  &-& \frac{1}{|\ka_i - \ka_{i+1}|^2} [ \ga^* \otimes \ga^* \;\ka_{i+1} \ka_i + 
  \ga \otimes \ga \;\ka^*_{i+1} \ka^*_i ] 
\eeqar
where $\otimes$ indicates that the two $\ga$-matrices are related to different femionic
lines in Fig.~3b. The first term on the r.h.s. of Eq.(\ref{VV}) contributes
to the exchange of quark and antiquark with equal
chiralities.
 We are
interested in the second term contributing to the opposite chirality exchange.

In  Regge asymptotics the transverse part of the exchanged fermion
propagator dominates i.e.
\beq{prop}
\frac{\hat{k}}{k^2} \longrightarrow \frac{1}{2}\big( \ga^* \frac{1}{\ka^*} +
\ga \frac{1}{\ka} \big) \;\;.
\eeq

A two particle amplitude $AB \rightarrow A'B'$ with a single fermion
exchange of definite chirality $A_{1F}(s,q)$ has the Regge asymptotic form
\beq{1f}
A_{1F}(s,q) = \Phi_{A'A}(q)\; \frac{s^{\al_F(q)}}{q}\; \Phi_{B'B}(q)
\eeq
where here and in the formulas below of this subsection $q$ is 
 the (transverse part of the) momentum transfer vector in the
complex number notation. The functions $\Phi_{A'A}(s,q)$
and $\Phi_{B'B}(s,q)$ are the impact factors  
describing the coupling of the Regge pole to the scattered
particles.

The Regge trajectory $\al_F(q)$ is given by
\beqar{traj}
 \al_F(q) &=& \frac{g^2 C_F}{(2\pi)^3} \;\bar{\al}_F(q)\;\;\;, 
\;\;\; C_F = \frac{N^2 - 1}{2N} \;\;,\nonumber \\
 \bar{\al}_F(q) &=& - \int \frac{d^2 \ka}{|q - \ka|^2}\cdot \frac{q}{\ka}
   \;\;\;,
\eeqar
$N$ being the number of colours.
Some intermediate regularization is needed to define the integral. It can be
omitted from the resulting equation (\ref{equation}) due to the 
cancellation of infrared
divergencies.

The Regge asymptotics of scattering amplitudes is represented in terms of
partial waves ${\cal A}^\pm(\om,q)$
\beqar{partwave}
A(s,q) &=&  A^+(s,q) + A^-(s,q) \nonumber \\
 A^\pm(s,q) &=& \int_{-i\infty}^{i\infty} 
\frac{d \om}{2\pi i} \;\ze^\pm(\om)
  \big(\frac{s}{\mu^2}\big)^\om \;{\cal A}^\pm(\om,q)
\eeqar
where $ A^+(s,q)$ ($ A^-(s,q)$) is the crossing even (odd) part of
the amplitude $A(s,q)$ and  
$\ze^\pm(\om)$ is the signature factor. For the leading behaviour of
the amplitude $T_\mu(p,q,\cs_\perp)$ related to $h_1(x,Q^2)$ (see 
Eqs.(\ref{IKh}) and (\ref{IKh1})) only small values of $\om$ are important.
Then in the case of positive signature the signature factor can be omitted, 
$\ze^+(\om) \approx 1$.

The partial wave describing the sum of the graphs in Fig.~2 has the form
\beq{definf}
 {\cal A}^+(\om,q) = \int d^2 \ka 
\frac{d^2 \bar{\ka}}{\bar{\ka}^2} \;\Phi^{(0)}(\ka,q)
\cdot f(\om,\ka,\bar{\ka},q) \cdot \Phi^{(P)}(\bar{\ka},q) \;\;.
\eeq
The impact factor $\Phi^{(0)}(\ka,q)$ describes the coupling to the external
currents ( see Eq.~(\ref{extcurrent})) 
and the impact factor 
$\Phi^{(P)}(\bar{\ka},q)$ describes the coupling of the scattered proton to the
exchanged reggeons. $f(\om,\ka,\bar{\ka},q)$ is the two-reggeon Green function
for which we would like to formulate now the equation describing the sum of
graphs in Fig.~2.

Writing the partial wave for  the one-fermion exchange amplitude (\ref{1f}) 
we see that  reggeized fermion exchange corresponds to the propagator
\beq{1fpartial}
  \frac{1}{\om - \al_F(q)} \cdot \frac{\ga}{q} + 
     \frac{1}{\om - \al^*_F(q)} \cdot \frac{\ga^*}{q^*} \;\;.
\eeq

The two-fermion exchange of opposite chirality is represented by
\beq{2fpartial}
\frac{1}{\om - \al_F(\ka) - \al_F( \ka - q)} \cdot \frac{\ga}{\ka} \otimes 
\frac{\ga}{\ka - q} + 
\frac{1}{\om - \al^*_F(\ka) - \al^*_F(\ka - q)} \cdot \frac{\ga^*}{\ka^*} 
\otimes  \frac{\ga^*}{\ka^* - q^*} \;\;.
\eeq

Now we have all buildings blocks (\ref{VV}), (\ref{2fpartial}) for the
equation describing the interacting two-fermion exchange. Choosing the
component in the reggeon-current coupling  proportional to
$\ga^*\ga_+$ (\ref{extcurrent}) (so the impact factor $\Phi^{(0)}(\ka,0)$ is
just 1)
we pick up the corresponding terms from 
 the propagators (\ref{2fpartial}) and
the interaction kernel (\ref{VV}). The resulting expression (proportional to 
 $\ga\ga_+$ ) implies that the 
 reggeon Green function
$f(\om,\ka,\bar{\ka})$ at $q = 0$ is the solution of the equation
\beq{equation}
\big[ \om - 2\al_F(\ka) \big] f(\om,\ka,\bar{\ka}) = \de^{(2)}(\ka -
\bar{\ka})
 + \frac{g^2C_F}{(2\pi)^3} \int d^2 \ka'\; \frac{2}{|\ka - \ka'|^2} \cdot
  \frac{\ka'}{\ka} \cdot f(\om,\ka',\bar{\ka}) \;\;\;.
\eeq
This equation is the analogon of the BFKL Pomeron equation \cite{BFKL} for
the case of opposite chirality fermion exchange. It has been considered
together with the equation for equal chirality fermion exchange in
\cite{K}.

The solution of equation (\ref{equation}) can be written in terms of the
eigenfunctions $\phi_{n,\nu}(\ka)$
\beq{phi}
\phi_{n,\nu}(\ka) = |\ka|^{2i\nu} \; \big(\frac{\ka}{|\ka|}\big)^{1 + n}
\eeq
and eigenvalues $\frac{g^2C_F}{8\pi^2}\;\Omega(n,\nu)$
\beq{eigenvalues}
\Omega(n,\nu) = 4\psi(1) - \psi(\frac{1}{2} + i\nu + \frac{n}{2}) - 
\psi(\frac{1}{2} - i\nu + \frac{n}{2}) - \psi(\frac{1}{2} + i\nu -
\frac{n}{2}) - \psi(\frac{1}{2} - i\nu - \frac{n}{2})
\eeq
of the homogeneous equation. We obtain
\beq{solution}
f(\om,\ka,\bar{\ka}) = \frac{1}{2\pi^2}\cdot \frac{1}{\ka^2}
\sum_{n = -\infty}^{\infty} \int_{- \infty}^{\infty} d\nu \cdot
\frac{\phi_{n,\nu}(\ka) \cdot \phi_{n,\nu}^*(\bar{\ka})\cdot
\big(\frac{\bar{\ka}}{\bar{\ka}^*}\big)}{\om -
\frac{g^2C_F}{8\pi^2}\;\Omega(n,\nu)} \;\;.
\eeq

After convolution with the impact factor $\Phi^{(0)}(\ka,0) = 1$ (see
Eq.(\ref{definf})) only the term with $n = 1$ contributes. The Regge singularity
appears at the value $\om_1$, where the two poles in $\nu$ from the
denominator in Eq.(\ref{solution}) pinch the integration contour
\beq{omega1}
\om_1 = \frac{g^2C_F}{8\pi^2}\;\Omega(1,0) = 0 \;\;.
\eeq 

Notice that for the singularity $\om_0$ of the contribution $n = 0$ we have
$\om_0 = \frac{g^2}{\pi^2}\;C_F\;\ln 2~>~\om_1$. The decoupling due to
the structure of $\Phi^{(0)}(\ka,0)$ prevents the $n = 0$ contribution from
dominating the asymptotics. This decoupling of the right-most singularity is
a special feature of the considered exchange channel.

\subsection{Coupling to the proton.}

The impact factor $\Phi^{(P)}(\bar{\ka})$ carries the non-trivial information
about the proton structure which we are not able to calculate from
perturbative QCD. We adopt the following form for this impact factor
\beq{impactP}
\Phi^{(P)}(\ka) = \frac{\hat{p}}{\sqrt{s}} \;(\cs_\perp k_\perp) \;
  \hat{k}_\perp \;\ga_5\;{\cal F}^{(P)}(|\ka|^2) \;\;\;.
\eeq

The function ${\cal F}^{(P)}(|\bar{\ka}|^2)$ can be a Gaussian with a width
determined by the scale of the proton mass.

With the ansatz (\ref{impactP}) we are able to calculate the partial wave
${\cal A}^+(\om,0)$ from Eq.(\ref{definf}). The inverse of Mellin 
transform (\ref{partwave}) 
gives the discontinuity  of the scattering amplitude
\beq{imagT}
{\rm Disc}\;T^*(p,q,\cs_\perp) = - i\cs^*\frac{1}
{\sqrt{\pi \Omega_0\;\ln (\frac{1}{x})}}\int^{Q^2} \frac{d^2
\ka}{|\ka|^2} \int d^2\bar{\ka} \; exp\big(-
\frac{\ln^2\frac{|\ka|^2}{|\bar{\ka}|^2}}{4\Omega_0 \ln (\frac{1}{x})} \big)
\cdot {\cal F}^{(P)}(|\bar{\ka}|^2)
\eeq
where
\beq{omega0}
\Omega_0 = \frac{g^2 C_F}{2\pi^2}\zeta(3) \;\;\;.
\eeq
The integral over $\ka$ can be expressed in terms of the incomplete gamma
function  $\Phi(x)$ $\cite{GR}$.
 We finally obtain the following expression for the leading small $x$
asymptotics of the structure function $h_1(x,Q^2)$
\beq{reggeh1}
h_1(x,Q^2)\Bigg|_{\rm Regge} = 
\frac{1}{2}\int d^2\bar{\ka}\; \Big( 1 + \Phi\Big(
\frac{\ln \frac{Q^2}{|\bar{\ka}|^2}}{\sqrt{4\Omega_0 \ln
(\frac{1}{x})}}\Big)\Big) \cdot {\cal F}^{(P)}(|\bar{\ka}|^2) \;\;\;.
\eeq
 Eqs.~(\ref{solution}),
(\ref{omega1}) and (\ref{reggeh1}) are our main results.

\section{Summing double-logarithmic contributions.}

\setcounter{equation}{0}
\subsection{The GLAP evolution of $h_1(x,Q^2)$.}

The contributions summed up above have been extracted in the leading $\ln s$
approximation, i.e. only the contribution of each loop with a logarithmic
longitudinal momentum integral has been taken into account. Now we consider
the contributions with a logarithm from the transverse momentum part of each
loop integral. It turns out that the longitudinal momentum integral in these
contributions can also be approximated in some region by a logarithm. Our
aim is to sum up the resulting double log contributions.

We start with the one loop graphs shown in Fig.~4 
which contribute to the $s$-channel discontinuity and we choose the axial
gauge $A_\mu q_1^\mu = 0$
\beq{1loop}
{\rm Disc}\; T^{(1)}_\mu(p,q,\cs_\perp) = \frac{i}{2}g^2C_F \int \frac{d^4
k}{(2\pi)^4}\; \frac{{\cal N}^{\rho\si}_{\mu} \cdot d_{\rho\si}(p - k)}
{(k^2 + i\epsilon)^2} \cdot (- 2\pi i)^2\; \de_+\Big[(p - k)^2\Big] \;
\de_+\Big[(k + q)^2\Big]
\eeq
where $d_{\rho\si}(k)$ is the nominator of the axial gauge gluonic
propagator and ${\cal N}^{\rho\si}_{\mu}$ is the matrix element related to
a quark line
\beqar{dN}
d_{\rho\si}(k) &=& g_{\rho\si} - \frac{q_{1\rho}k_\si + k_\rho q_{1\si}}
{q_1\cdot k} \nonumber \\
{\cal N}^{\rho\si}_{\mu} &=& \bar{u}(p,\cs_\perp) \ga^\rho \hat{k}
\Big[ \ga_5 \ga_\mu (\hat{k} + \hat{q}) + (\hat{k} + \hat{q})\ga_5 \ga_\mu
\Big] \hat{k} \ga^\si u(p,\cs_\perp) \;\;\;.
\eeqar
We restrict ourselves to  vanishing momentum transfer. The scattering quark
with momentum $p$ is assumed to be on mass-shell and we project out the
contribution proportional to the transverse polarization 
vector $\cs_\perp$ as in
section 2.1
\beq{nomin} 
{\cal N}^{\rho\si}_{\mu}d_{\rho\si}(p - k) = - \frac{4 s k^2}{q_1\cdot (p - k)}
\Bigg[ 2\cs_{\perp\mu} (q_1 k) + k^2 \cs_{\perp\mu} - k_\mu (\cs_\perp
k)\Bigg] \;\;\;.
\eeq
We parametrize the loop momentum $k$ as $k = \al\;q_1 + \be\;p + k_\perp$
and we use the complex number notation (\ref{compnot}). 
The contribution with a
logarithmic $\ka$ integral arrises from the first term in the square bracket
of (\ref{nomin})
\beqar{calcsplit}
&{\rm Disc}&\; T^{(1)}_\mu(p,q,\cs_\perp) \nonumber \\ 
&=& \cs_{\perp\mu}\frac{(- i)g^2C_F}{2\pi}
\int d\al \; d\be\; \frac{d|\ka|^2}{|\ka|^2}\cdot \frac{\be}{1 - \be}\;
\de_+\Bigg[\be - x - \frac{|\ka|^2}{s}\cdot \frac{1 - x}{1 - \be}\Bigg]\;
\de_+\Bigg[\al + \frac{|\ka|^2}{s(1 - \be)}\Bigg] \nonumber \\
&\approx & \cs_{\perp\mu}\frac{(- i)g^2C_F}{2\pi}\cdot \frac{x}{1 - x}\int
\frac{d|\ka|^2}{|\ka|^2} \;\;\;.
\eeqar
As a contribution to deep-inelastic scattering the $\ka$ integral results
into $\ln Q^2$ and the coefficient is to be identified (see
Eq.(\ref{IKh1})) 
as the parton (GLAP)
splitting function (before regularizing the singularity at $1 - x$
appropriately)
\beq{splittingf}
   P^{(0)} (x) = 2C_F\;\frac{x}{1-x}
\eeq
We recover the known one-loop result for the splitting function 
\cite{ArtruMekhfi} respectively the resulting  anomalous dimensions \cite{BuFKL}.

We compare the result (\ref{calcsplit}) 
from the viewpoint of Regge asymptotics with the one obtained
in section 2. The contribution (\ref{calcsplit}) is down by one power of $s$
compared to the ones analysed in (\ref{reggeh1}). From this we learn that
  for $h_1(x,Q^2)$ 
the small $x$ asymptotics of the GLAP evolution does not reproduce the
leading perturbative Regge behaviour.

\subsection{Double-logs in higher orders.}

Now we analyse how the contributions with a logarithmic $\ka$ integral
emerge in higher loops. In particular we are interested in the integration
region resulting in a logarithmic contribution from each transverse momentum
loop integral.

The GLAP equation, describing the large $Q^2$ behaviour of structure
functions, sums the leading log contributions of the considered type only
for  strongly ordered transverse momenta. We shall see that in our
case this is not the only region which gives  leading logarithms from 
the transverse momentum integrals.

We analyse the two-loop integral from Fig.~5. We extract
from the numerator, consisting of the Dirac trace and the 
factors $d_{\mu\nu}$ of
the gluons propagators, the contributions proportional to $k_1^2\cdot k_2^2$
resulting in the leading logarithms. We obtain
\beqar{2loop}
{\rm Disc}\;T^{(2)}_\mu(p,q,\cs_\perp) &=& \cs_{\perp\mu}\; 
\frac{(- i)g^4s^3}{(2\pi)^5}\; \int d\al_1\; d\al_2\; d\be_1\; d\be_2\;
\frac{d^2\ka_1}{s\al_1\be_1 - |\ka_1|^2}\; \frac{d^2\ka_2}{s\al_2\be_2 -
|\ka_2|^2} \nonumber \\ 
&\cdot&P^{(0)}(\be_2) P^{(0)}(\frac{\be_1}{\be_2}) 
\cdot \de_+\Big[s(1 + \al_1)(\be_1 - x) - |\ka_1|^2\Big] \nonumber \\
&\cdot&\de_+\Big[s(\al_1 - \al_2)(\be_1 - \be_2) - |\ka_1 - \ka_2|^2\Big]\;
\de_+\Big[s(\al_2(\be_2 - 1) - |\ka_2|^2\Big] \;.
\eeqar

We perform the integration over $\al_1$ and $\al_2$ and obtain that both
$\ka$ integrals are logarithmic in the region
\beqar{intregion}
&1& \gg \be_2 \gg \be_1 \gg \frac{\mu^2}{s} \nonumber \\
&\mbox{}& |\ka_2^2| < |\ka_1^2|\;\frac{\be_2}{\be_1}
\eeqar
( $\mu^2$ is the infrared cut-off).
This is the integration range encountered in the leading double-log
asymptotics of equal chirality  fermion exchange \cite{Gorshkov}. Because of
this analogy it is not necessary to go beyond two-loops. We are able to
write down the integral equation summing up all contributions logarithmic in
both the $\ka$ and $\be$ integrals.

It is convenient to write the equation for the sum of ladders without the
external currents as denoted in Fig.~2 by $f$ (there $f$ refers to the
partial wave, the corresponding amplitude will be denoted by  
 $A(\be,|\ka^2|)$)
\beqar{eqA}
A(\be_1, |\ka_1^2|) &=& \frac{g^2}{2(2\pi)^2}\;
       \int_{\be_1}^{1} \frac{d\be_2}{\be_2}\; 
   \int_{\mu^2}^{|\ka_1^2|\frac{\be_2}{\be_1}} \frac{d|\ka_2^2|}{|\ka_2^2|}\;
  P^{(0)}\Bigg(\frac{\be_1}{\be_2}\Bigg)\; A(\be_2, |\ka_2^2|) \;\;+
  \;\; A_0(\be_1,  |\ka_1^2|)  \nonumber \\
 A_0(\be_1,  |\ka_1^2|) &=& \frac{g^2}{2(2\pi)^2}\;  P^{(0)}(\be_1) \;\;.
\eeqar

The structure function $h_1(x,Q^2)$ in the small $x$ region
($\mu^2\ll Q^2 \ll s$) is obtained from $A(\be_1,|\ka_1^2|)$
by
\beq{hA}
h_1(x, Q^2) = \int_{\frac{\mu^2}{s}}^1 d\be_1\; \int_{\mu^2}^{s/4}
\frac{d|\ka_1^2|}{|\ka_1^2|}\; \de\Bigg[ \be_1 - x -
\frac{|\ka_1^2|}{s}\Bigg]\; A(\be_1, |\ka_1^2|)
\eeq

Within the double-log approximation we rewrite the equation
(\ref{eqA}) in terms of $\tilde{A}~=~\be_1^{-1}~A$ and approximate the
splitting function $P^{(0)}(\be)$ (\ref{splittingf}) by
\beq{splittregge}
  P^{(0)}(\be) \approx 2C_F\;\be \;\;\;.
\eeq
The resulting double-log equations read
\beqar{eqAtilde}
\tilde{A}(\be_1, |\ka_1^2|) &=& \frac{g^2C_F}{(2\pi)^2}\;
       \int_{\be_1}^{1} \frac{d\be_2}{\be_2}\; 
   \int_{\mu^2}^{|\ka_1|^2\frac{\be_2}{\be_1}} \frac{d|\ka_2^2|}{|\ka_2^2|}\;
  \tilde{A}(\be_2, |\ka_2^2|) \;\;+
  \;\; \tilde{A}_0(\be_1,  |\ka_1^2|)  \nonumber \\
 \tilde{A}_0(\be_1,  |\ka_1^2|) &=& \frac{g^2C_F}{(2\pi)^2}
\eeqar
and
\beq{hAtilde}
h_1(x, Q^2) = \int_{\frac{\mu^2}{s}}^1 d\be_1\;\be_1\; \int_{\mu^2}^{s/4}
\frac{d|\ka_1^2|}{|\ka_1^2|}\; \de\Bigg[ \be_1 - x -
\frac{|\ka_1^2|}{s}\Bigg]\; \tilde{A}(\be_1, |\ka_1^2|) \;\;.
\eeq

Eq.(\ref{eqAtilde}) is to be compared with 
the corresponding equation for the flavour
non-singlet structure function $F_1(x,Q^2)$ \cite{EMR}. The solution can be
written in terms of a Bessel function $I_2(z)$ \cite{Gorshkov}.

\subsection{Separation of the smallest transverse momentum.}
 
The double-log contribution can also be treated by the method of separation
of softest particle \cite{KL}. It leads directly to simpler equations in
terms of partial waves with a quadratic term. A particular advantage of this
approach is that both signature parts of the amplitude $A(s,\ka^2)$ can be
calculated in the double-log approximation
\beqar{partf}
A(s,\ka^2) &=&  A^+(s,\ka^2) + A^-(s,\ka^2) \nonumber \\
 A^\pm(s,\ka^2) &=& \int_{-i\infty}^{i\infty} 
\frac{d \om}{2\pi i} \;\ze^\pm(\om)
  \big(\frac{s}{|\ka^2|}\big)^{- 1 + \om} \;f^\pm(\om) \;\;\;.
\eeqar
Note that because 
 this double-log contribution behaves like $s^{-1}$, the definition of the
partial wave (\ref{partf}) differs in comparision with Eq.(\ref{partwave}). 
For small $\om$ the signature factor behaves like 
\beqar{sign}
\ze^\sigma(\om) \approx \left\{ \begin{array}{r@{\quad \quad}l}
 - \frac{i\pi}{2}\om & \si = +1 \\
1 & \si = -1
\end{array}\right.\;\;\;.
\eeqar

There are two types of double-log contributions from loops with the smallest
transverse momentum. The contribution of loops with a soft gluon of
bremsstrahlung type is expressed in terms of the original amplitude with the
gluon attached to the external lines.
The contribution of a two-particle intermediate state with the smallest
$\ka$ is expressed in terms of a loop integral where the amplitude appears
twice. The latter contribution results in a quadratic term in 
equations for $f^\pm(\om)$.
The double-log integration range established in (\ref{intregion}) implies
the existence of this contribution of the soft two-particle intermediate
state in our case. In the previous subsection we summed in fact the
double-log contributions of the latter type. The bremsstrahlung type
contribution cancel in fact in the colour singlet 
channel of negative signature \cite{KL}.
The integral equation (\ref{eqA}) applies just to this case.

We need the result for the partial waves for colour singlet exchange. It is
given in \cite{KL} ( in our case $a_0 = 2C_F$ and the roles of positive and
negative signatures are interchanged ),
\beqar{solf}
f_0^-(\om) &=& 4\pi^2\;\om\;\Bigg( 1 - \sqrt{1 - \frac{g^2C_F}{\pi^2
\om^2}}\;\;\Bigg) \;\;\;,  \nonumber \\
f_0^+(\om) &=& 4\pi^2\;\om\;\Bigg(1 - 
\sqrt{ 1 - \frac{g^2C_F}{\pi^2 \om^2}\Big( 1 - \frac{1}{2\pi^2\om}\cdot
f_V^-(\om)\Big)}\;\;\Bigg) \;\;\;.
\eeqar

In Eq.(\ref{solf}), the positive signature partial wave $f_0^+(\om)$ 
is expressed in terms of 
the negative signature partial wave of the
colour octet channel  $f_V^-(\om)$ equal \cite{KL}   
\beq{octet}
f_V^-(\om) = 2g^2  N \frac{d}{d\om}\;\ln \;\bigg\{ e^{\frac{\om^2}{4\bar{\om}^2}} {\cal
D}_p(\frac{\om}{\bar{\om}})\bigg\} \;\;,\;\;\;\bar{\om} =
\Bigg(\frac{g^2N}{4\pi^2}\Bigg)^\frac{1}{2}
\eeq
where ${\cal D}_p(z)$ is the parabolic cylinder function 
\cite{GR} with $p = -
\frac{1}{2N^2}$. Eq.(\ref{solf}) implies that the leading Regge
singularity for positive signature lies to the right 
of the negative signature singularity by a small amount
\beq{shift}
\om_0^+ \approx \om_0^- \Bigg(1 + \frac{1}{2N^2}\Bigg) \;\;,\;\;\;\om_0^- =
\Bigg(\frac{g^2C_F}{\pi^2}\Bigg)^\frac{1}{2}
\eeq
The partial waves (\ref{solf}) and (\ref{octet}) determine
 the Regge asymptotics of the scattering amplitude 
in the double-log approximation.
In particular, 
the result (\ref{shift}) for the leading singularity $\om_0^+$
implies the existence of a contribution to $h_1(x,Q^2)$ with
\beq{dlog}
h_1(x,Q^2)\Bigg|_{{\rm double-log}} 
\sim \Bigg(\frac{1}{x}\Bigg)^{- 1 + \om_0^+}\;\;\;.
\eeq
The large $Q^2$ behaviour in the small $x$ region can also be extracted from
the partial waves (\ref{solf}) and (\ref{octet}). Indeed they determine the
asymptotic anomalous dimensions $\nu(j)$ with the contributions proportional
to $\Big(\frac{g^2}{(j + 1)^2}\Big)^n$ summed to all orders $n$
\beq{anomdim}
\nu(j) = \frac{1}{8\pi^2} \cdot f_0^+(j + 1)\;\;\;,\;\;\;\;\om = j + 1
\;\;\;.
\eeq
We checked that the first order term coincides with the leading term for 
$\om  \rightarrow 0$ of the moments calculated from the one-loop
splitting function (\ref{splittingf})
\beq{anom1loop}
\nu^{(1)}(j) = \frac{g^2}{8\pi^2} \int_0^1 dx\; x^{j - 1}\cdot
P^{(0)}(x)\Bigg|_{j \rightarrow - 1} = \frac{g^2}{8\pi^2}\cdot \frac{2C_F}{j
+ 1}\;\;\;.
\eeq

\section{Discussion}

There are arguments in favour of the similarity of 
the transversity distribution
$h_1(x,Q^2)$ and the helicity distribution $g_1(x,Q^2)$ at $x \rightarrow 1$
 \cite{JaffeErice}. In particular the one-loop parton splitting function
for the flavour non-singlet $g_1(x,Q^2)$ and for $h_1(x,Q^2)$ match at 
$x \rightarrow 1$.

Summarizing our results we have seen, however,  that at small $x$ 
the structure function $h_1(x,Q^2)$ behaves
 quite different  from 
$g_1(x,Q^2)$. For $g_1(x,Q^2)$ 
the flavour singlet part  leads to a rapid
increase towards small $x$, and  also the non-singlet part shows some increase 
\cite{BER}.

Instead for  $h_1(x,Q^2)$
at small $x$ we have encountered two contributions.

The first contribution is roughly constant in $x$ and  shows also a weak 
$Q^2$ dependence. It results from the sum of the leading perturbative Regge
contributions from the exchange of two reggeized fermions with opposite
chirality. Contrary to the case of equal chirality they are of simple
logarithmic type with the logarithm resulting from the longitudinal momentum
integration in each loop. Adopting a model for the corresponding proton
impact factor we have given a rather explicite form of this first
contribution, Eq.(\ref{reggeh1}).

The second contribution is non-leading compared to the first one at small
$x$, Eq.(\ref{dlog}). It arises as a sum of all contributions with leading
logarithms both from longitudinal and transverse momentum integrations. It
includes the small $x$ asymptotics of  GLAP evolution, which corresponds
to restricting the transverse momentum integration range to the strongly
ordered configurations. Since the double log range exceeds  this strongly
ordered one, the actual small $x$ behaviour of this second contribution
deviates from the GLAP behaviour in the same sense as in the flavour
non-singlet $F_1(x,Q^2)$ or $g_1(x,Q^2)$ cases \cite{EMR}, \cite{BER} (see
also \cite{BlVogt}).
Compared to the structure functions mentioned above it appears as an
unusual feature here that the dominant contribution 
at small $x$ does not include  GLAP
asymptotics. This fact makes  the studies of $h_1(x,Q^2)$ 
in the region of $x$ in which both contributions play a role very
interesting.

We have concentrated on the contribution with positive signature. The
leading $\ln{s}$ asymptotics in the negative signature case is
influenced by a three Reggeon exchange, for which no explicit results are
known.

The results derived in this paper are obtained within the leading 
logarithm approximation and the double logarithmic approximation. Due to that 
we are unable to give prediction about the relative weight of two
contributions mentioned above and to give an estimate at which value of $x$
a deviation from GLAP evolution has to be expected.

\vspace*{1cm}
\noindent{\Large \bf  Acknowledgements.}

\vspace*{.5cm}

\noindent A.Sch.'s and L.Sz.'s research was supported by the DFG grant (Sch 458/3).

\noindent L.M.'s research was supported by BMBF and by KBN grant 2~P03B~065~10.

\noindent R.K. and L.Sz. acknowledge the  support of the German-Polish agreement on
scientific and technological cooperation N-115-95.
 
\noindent L.Sz. would like to acknowledge the warm hospitality extended to
him at J.W.~Goethe Universit\"at Frankfurt am Main.

\vspace*{1.5cm}
\noindent{\Large \bf Figures captions.}
 \newline
\newline
{\bf Fig.~1:} Lowest order contribution to $T_\mu$. The wavy line, the dotted
line and the solid line denote the scattered axial meson, the scattered
scalar and the exchanged quark, respectively. 
\newline
\newline
{\bf Fig.~2:} The effective ladder diagram. The bold-solid line denotes the
exchanged reggeized quark. The dashed line denotes produced gluon. The
arrows on both sides of reggeized quarks emphasize that they have the same
helicities (opposite chiralities). The bold dot denotes the effective 
vertex for gluon production. The dashed-dotted line emphasizes the
calculations of $s$-channel discontinuity. 
\newline
\newline
{\bf Fig.~3:} The effective vertex for gluon production and the interaction
kernel.
\newline
\newline
{\bf Fig.~4:} The one-loop contribution to $T_\mu$.
\newline
\newline
{\bf Fig.~5:}  The two-loop contribution to $T_\mu$.

\newpage

\begin{center}
\leavevmode
\epsfxsize=16.0cm
\epsfbox{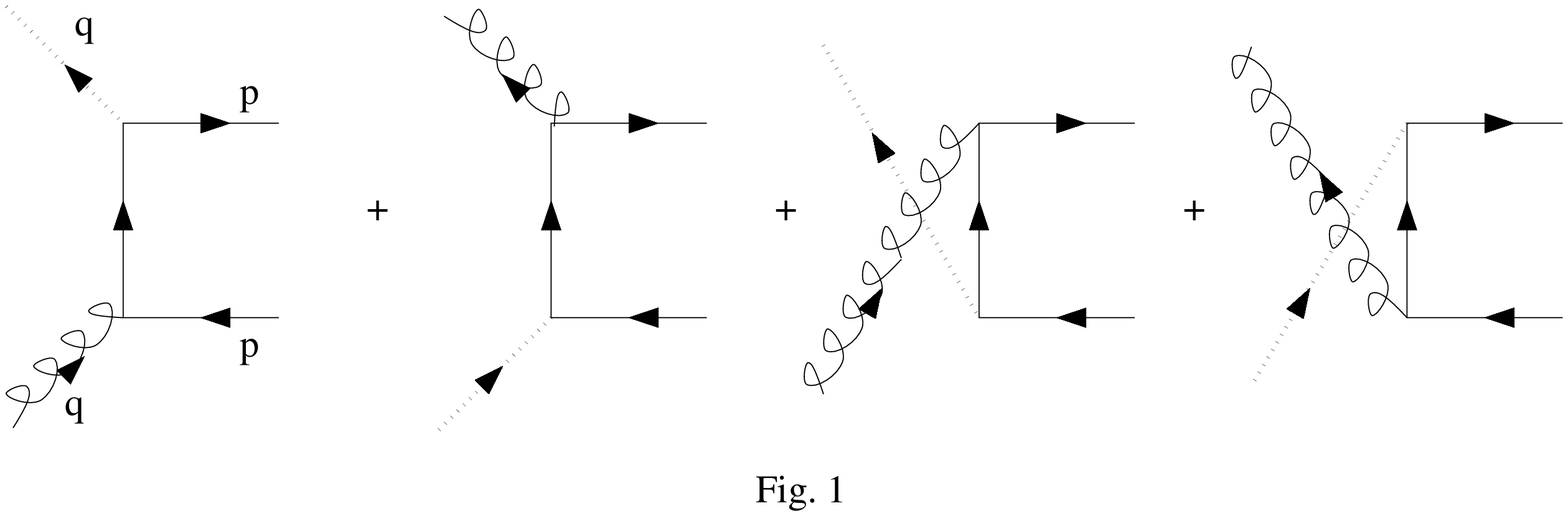}
\newline
\newline
\end{center}
\begin{center}
\leavevmode
\epsfxsize=16.0cm
\epsfbox{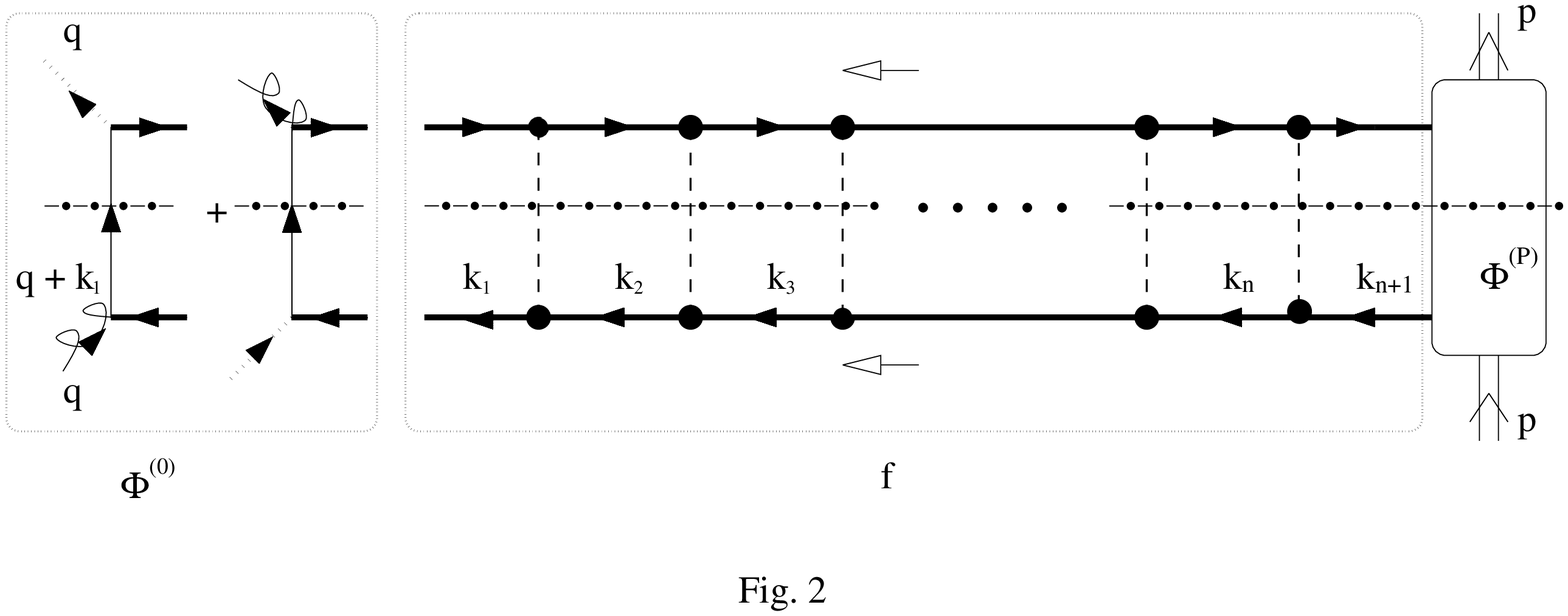}
\newline
\newline
\end{center}
\begin{center}
\leavevmode
\epsfxsize=16.0cm
\epsfbox{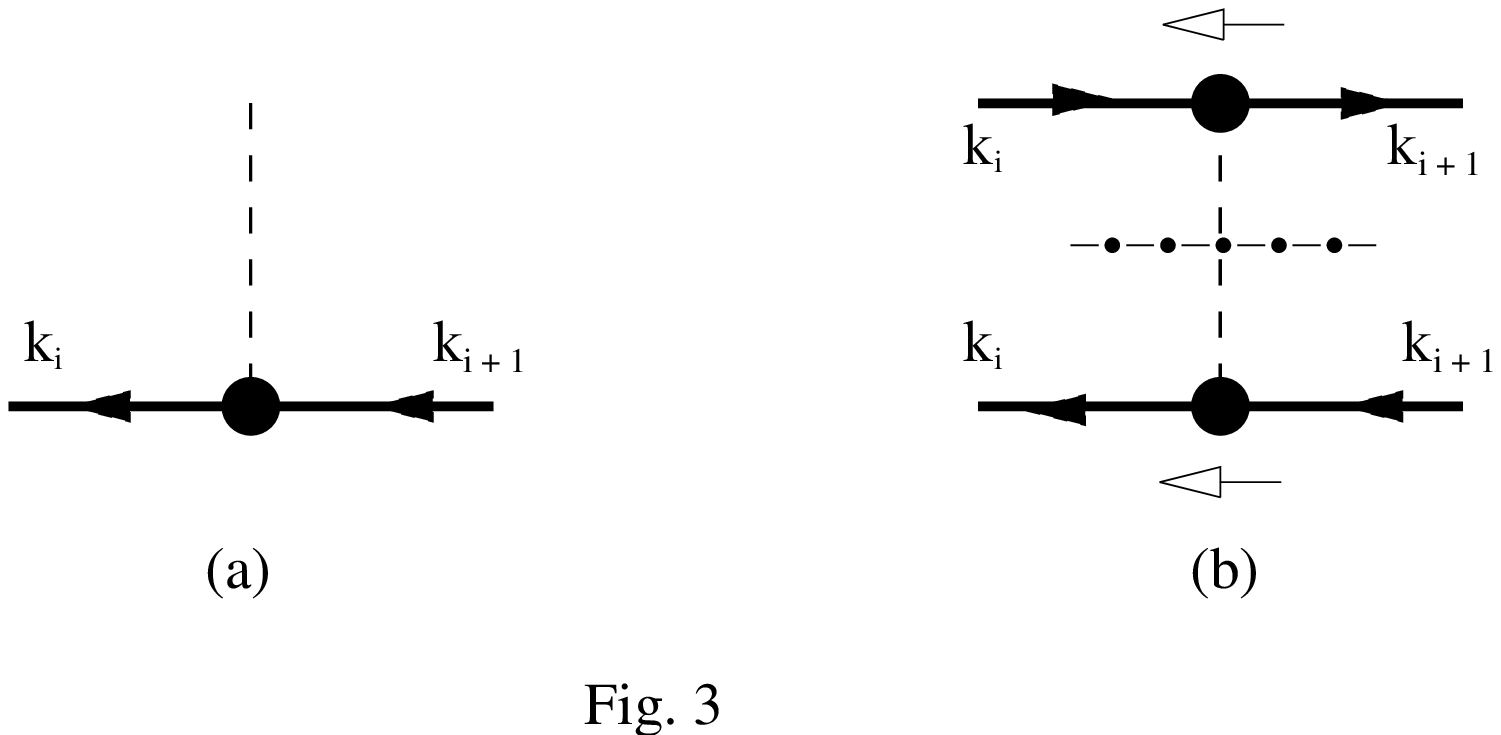}
\newline
\newline
\end{center}
\newpage   
\begin{center}
\leavevmode   
\epsfxsize=16.0cm
\epsfbox{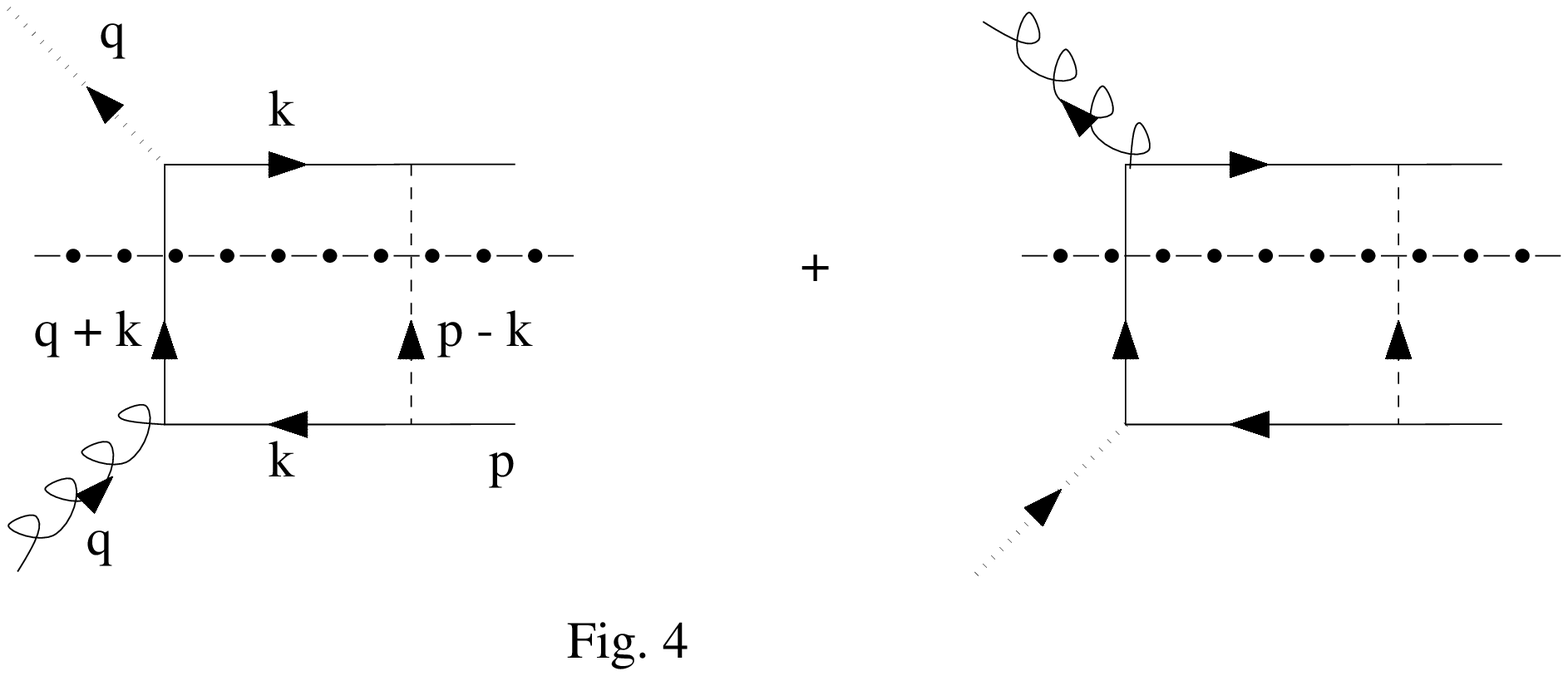}
\newline
\newline
\end{center}
\begin{center}
\leavevmode   
\epsfxsize=16.0cm
\epsfbox{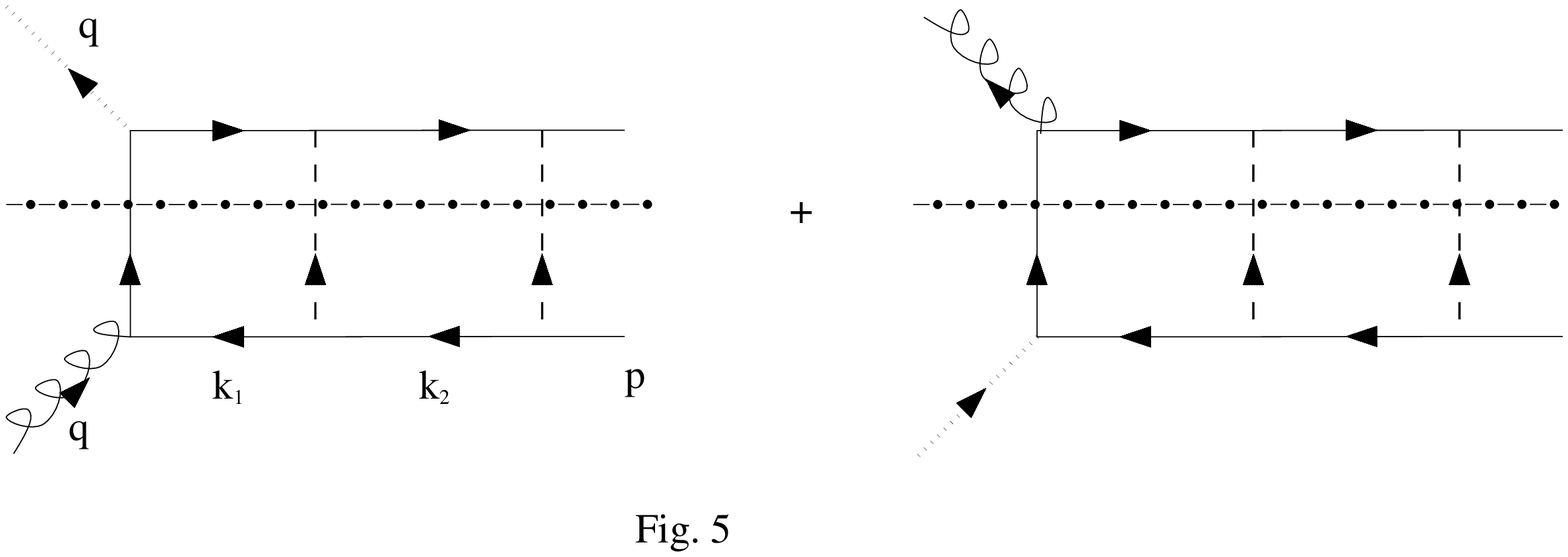}
\newline
\newline
\end{center}

\end{document}